# Analyzing Multi-Factor Authentication Through Cryptographic Security Properties

Ryan Tipping
School of Computing
Montclair State University
Montclair, New Jersey
tippingr1@montclair.edu

Yousef Tahboub
School of Computing
Montclair State University
Montclair, New Jersey
tahbouby1@montclair.edu

Krishna Bodige
School of Computing
Montclair State University
Montclair, New Jersey
bodigek1@montclair.edu

***Abstract***— Modern authentication systems use cybersecurity techniques to validate the identity of the person (or applications acting on behalf of the person) as a primary defense against unauthorized access. While initially built around single mechanisms such as usernames/passwords, physical tokens, or biometrics, current systems have evolved into multi-factor authentication (MFA) platforms that combine multiple mechanisms. Among them, several focus on strategies that prevent replay attacks (i.e. the reuse of a component that could have been potentially compromised).

This study explores real-world MFA systems, focusing specifically on time-based-one-time passwords (TOTP) and push-based authentication. These authentication methods are designed to provide an additional layer of security; our motivation stems from the need to explore whether real-world MFA implementations can still exhibit weaknesses despite relying on strong cryptographic foundations. This study examines how key security properties such as freshness, session binding, and server-side state are implemented in real systems. The systems were tested to analyze performance in normal and misconfigured conditions to identify potential weaknesses. These properties are essential for ensuring that authentication requests are valid and cannot be reused or manipulated. If these properties are not properly enforced, the system may fail to provide the level of security it is intended to guarantee. The results show that many practical weaknesses in MFA systems stem from failures in enforcing protocol-level guarantees in real deployments. This topic is especially important because MFA is widely trusted, yet even small implementation issues can lead to serious vulnerabilities in real-world applications.

## I. INTRODUCTION AND MOTIVATION

Modern authentication systems use cryptographic techniques to ensure properties like confidentiality, integrity, and authentication. MFA improves security by requiring multiple forms of verification, such as time-based one-time passwords (TOTP) and push-based authentication. TOTP systems are built using hash-based message authentication (HMAC), while push-based rely on digital signatures and secure communication channels.

Although these methods are built on strong cryptographic foundations, the security of MFA depends on more than the underlying algorithm. Real-world implementations must properly enforce assumptions such as freshness (token valid for a short time), session binding (authentication is tied to a specific session), and state consistency (the server correctly tracks the authentication process). These properties are critical for preventing attacks like replay, token reuse, and unauthorized approvals. However, in practice, these assumptions are not always enforced correctly. As a result, vulnerabilities can emerge even when the underlying cryptographic methods remain secure. Small implementation mistakes or infrastructure weaknesses can reduce the effectiveness of secure authentication systems.

This project examines the gap between theoretical security and real-world implementation by analyzing how MFA systems behave under different conditions. Specifically, it evaluates TOTP and push-based MFA systems to identify where enforcement of freshness, session binding, and server-side state breaks down, and how those failures impact the overall security of the system.

## II. RELATED WORK

Previous research has studied how secure multi-factor authentication (MFA) systems are in real-world use.

Reynolds et al. analyzed authentication logs from universities and found that while MFA does improve security, it also introduces usability issues with more than 1 in 20 authentication attempts failing [1]. Their findings show that system-level factors like timeouts, user behavior, and session handling can affect how MFA performs in practice.

Bonneau et al. developed a framework for evaluating authentication systems and analyzing many authentication methods, including OTP-based systems [2]. Their work showed that no authentication system achieves all desired security, usability, and deployability properties at once. This highlights how real-world design constraints influence the effectiveness of security mechanisms.

Mahdad et al. demonstrated attacks against mobile notification-based authentication and identified protocol-level weaknesses in push-based MFA despite the use of strong cryptographic mechanisms [3]. Their work shows that attackers can exploit implementation and user interaction weaknesses to manipulate users by approving malicious login requests.

Building on these studies, this project focuses on how MFA systems enforce core security properties in practice. While prior work has identified usability limitations and protocol-level attacks, this study extends that research by directly testing MFA implementations to evaluate how well they enforce freshness, session binding, and server-side state management.

## III. METHODOLOGY

This project implemented a simulated MFA system to evaluate common weaknesses in its deployment. The system supported two authentication methods: time-based one-time passwords and push-based authentication approvals. These methods were selected because they are commonly used in modern authentication systems and are frequent targets in real-world attack scenarios.

The TOTP-based system architecture is shown in Fig. 1. The local machine included the login interface, TOTP generation script, testing scripts, and a local credential file. The remote VPS hosted the Flask server behind an Nginx reverse proxy. The Flask application handled login, OTP verification, replay protection using an in-memory used_tokens guard, and credential validation through users.json.

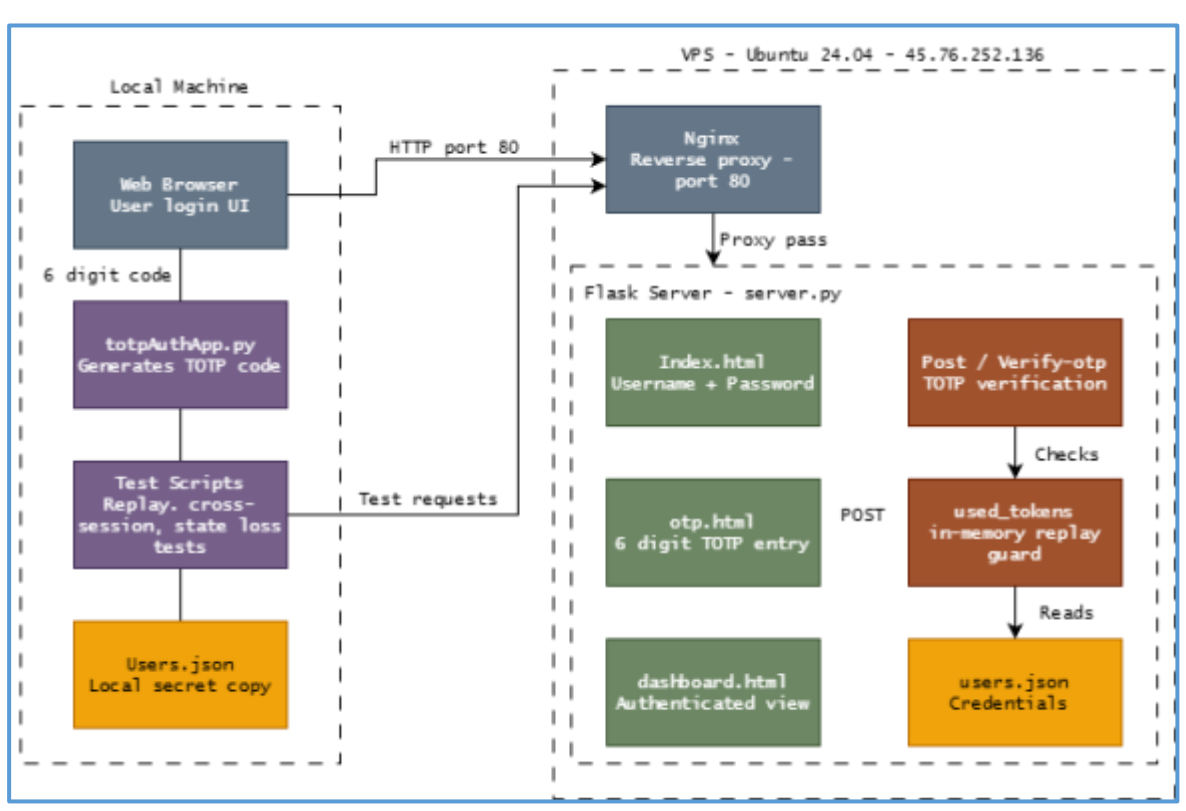

*Fig. 1. TOTP system overview showing the local test environment, VPS deployment, Flask server components, Nginx reverse proxy, OTP verification flow, and replay/state-loss testing components.*

The push-based system architecture is shown in Fig. 2. Like the TOTP setup, the system used a local machine to simulate user interaction and testing behavior. The remote VPS hosted the Flask application and Nginx reverse proxy. The push-based system included endpoints for login, MFA approval or denial, and MFA status checking. The system also included pages for user login, push status polling, authenticated access, and account registration. Test scripts were used to evaluate replay behavior, cross-session approval reuse, and state-loss conditions.

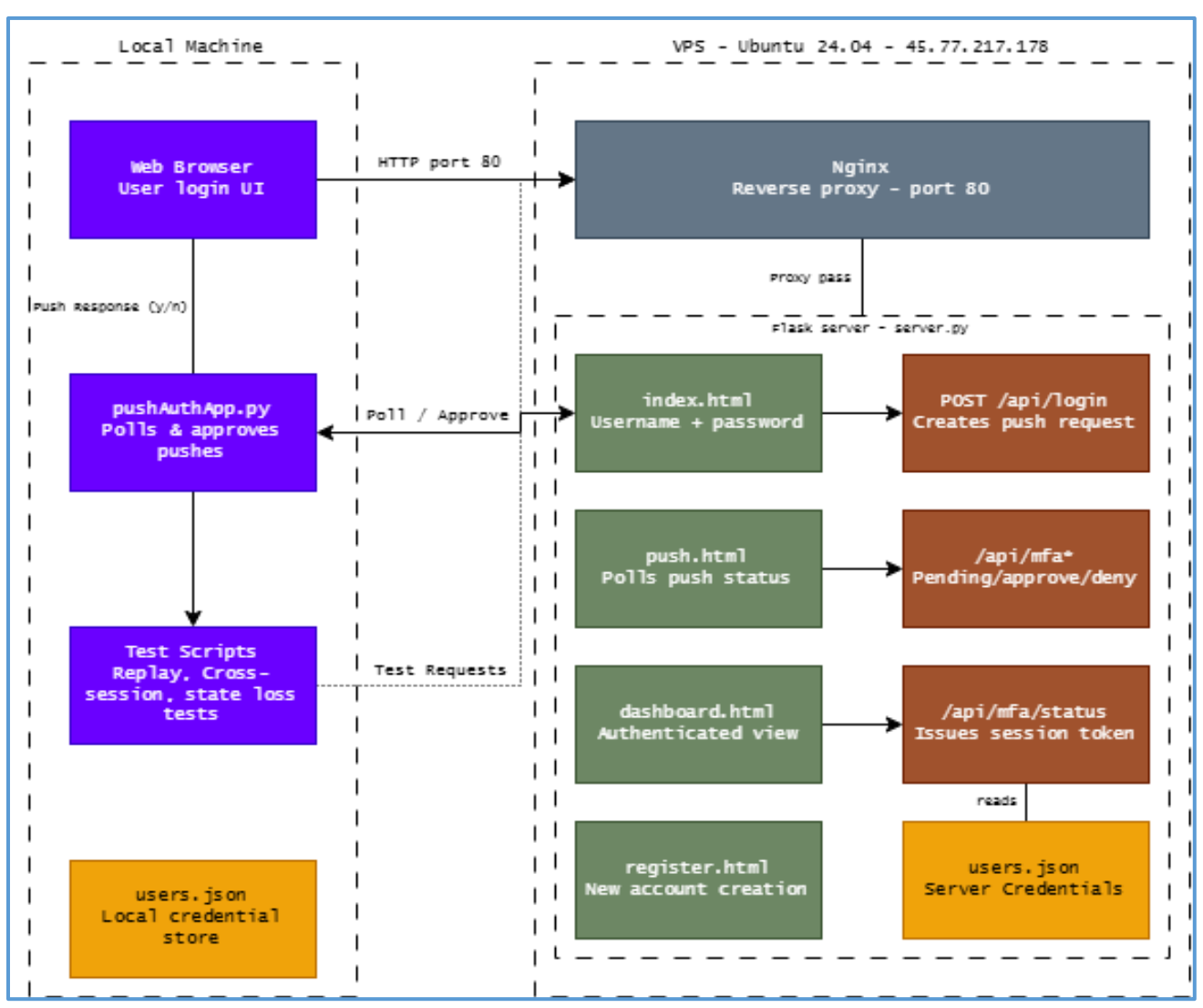

*Fig. 2. Push-based system overview showing the local test environment, VPS deployment, Flask server components, push request creation, approval/denial handling, MFA status checking, and session token issuance.*

The development process began with the creation of a simple front-end environment that simulated a standard login exchange. After the initial credential submission, the user was required to complete an additional authentication step before access was granted. The system was first developed and tested in a local environment to verify the authentication flow, API behavior, and MFA validation process. After local testing was completed, the system was deployed using two separate virtual private servers. One VPS hosted the TOTP-based authentication system, while the second VPS hosted the push-based authentication system. This separation allowed each MFA method to be tested independently while still representing a realistic remote deployment environment.

The implementation used Flask to build the server-side application and API endpoints for login handling, OTP verification, and push approval processing. The 'pyotp' library was used to generate and verify time-based one-time passwords. Standard cryptographic libraries were also used to support secure authentication logic and maintain consistency with common security practices.

The testing phase focused on three main security properties: freshness, session binding, and state management. Freshness testing evaluated whether authentication codes and approval requests remained valid only for a short period of time. Different validity windows were tested to observe how longer expiration times could increase the opportunity for an attacker to reuse or submit an intercepted MFA response.

Session binding testing examined whether each MFA code or push approval was linked to a specific login attempt. This test was used to determine whether an authentication response could be reused across multiple sessions. Proper session binding is necessary to prevent an MFA response from being transferred from one login attempt to another.

State management testing evaluated how the system handled used tokens, expired codes, and completed approvals. A secure MFA system must track whether an authentication response has already been used and reject repeated submissions. This portion of testing demonstrated how weak or missing server-side state tracking could create opportunities for replay attacks or unauthorized access.

Overall, the methodology combined local development, remote VPS deployment, and targeted security testing to evaluate how MFA systems may fail when freshness, session binding, and state management are not properly enforced. The use of separate VPS environments for OTP and push-based authentication allowed both methods to be analyzed individually under realistic deployment conditions.

## IV. EXPERIMENTAL RESULTS AND DISCUSSION

Three experiments were conducted to evaluate the enforcement of freshness, session binding, and server-side state management across both the TOTP-based and push-based MFA systems. Each test was designed to isolate a specific security property and observe how the system behaved under normal and intentionally weakened configurations.

### A. Test 1: Progressive Replay Window Expansion

The first experiment measured how the size of the validation window affected the feasibility of replay attacks. For the TOTP system, the allowed time window was progressively expanded from the default ±1 window (approximately 60 seconds) to ±2 and then ±4 windows. For the push-based system, the approval expiry duration was increased from 5 seconds to 30 seconds and then to 120 seconds. At each setting, previously captured codes and approval responses were replayed at increasing delays to determine when acceptance was possible.

**TABLE I**

*TOTP Replay Window Expansion Results*

| Window | Immediate | After 30s | After 60s | After 120s |
|---|---|---|---|---|
| **±1 (default)** | Pass | Pass | Fail | Fail |
| **±2 windows** | Pass | Pass | Pass | Fail |
| **±4 windows** | Pass | Pass | Pass | Pass |

*Note. This table shows TOTP acceptance under progressively larger replay windows.*

**TABLE II**

*Push MFA Replay Window Expansion Results*

| Expiry | Immediate | After 30s | After 58s | After 70s |
|---|---|---|---|---|
| **5 sec** | Pass | Fail | Fail | Fail |
| **30 sec** | Pass | Pass | Fail | Fail |
| **120 sec** | Pass | Pass | Pass | Pass |

*Note. This table shows push MFA acceptance under progressively larger approval expiry windows.*

The results demonstrate that window size directly controls the replay attack surface. At the default TOTP setting of ±1 window (~60 seconds total), replayed tokens were rejected after 60 seconds, making real-world exploitation impractical given the difficulty of intercepting and reusing a token within that window. However, at ±4 windows (~150 seconds), all replayed tokens were accepted regardless of delay, making man-in-the-middle replay attacks genuinely feasible. A similar pattern was shown in the push-based system. The 5-second expiry was highly restrictive, while the 120-second window provided an attacker with enough time to redirect and reuse an approval. Both TOTP ±1 and push with a 30-second expiry represent equivalent default security postures, with replay risk increasing greatly as window sizes are increased.

### B. Test 2: Cross-Session Confusion Under Load

The second experiment checked whether authentication responses were properly bound to a specific login session. Ten concurrent login sessions were opened simultaneously, and a single OTP or push approval was generated for each. In the TOTP test, the same code was submitted across all ten sessions. In the push test, each session generated a distinct push request with a unique identifier.

**TABLE III**

*TOTP Cross-Session Results (10 Concurrent Sessions)*

| | Without Replay Prevention | With Replay Prevention |
|---|---|---|
| **Sessions tested** | 10 | 10 |
| **Successes** | 10 | 1 |
| **Denials** | 0 | 9 |

*Note. This table compares TOTP behavior under concurrent sessions with and without replay prevention.*

**TABLE IV**

*Push MFA Cross-Session Results (10 Concurrent Sessions)*

| Metric | Result |
|---|---|
| **Sessions tested** | 10 |
| **Unique push IDs** | 10 / 10 |
| **Approved sessions** | 10 / 10 |
| **Collisions / errors** | 0 |

*Note. This table shows push MFA behavior under ten concurrent login sessions.*

Without prevention, the TOTP system accepted the same token across all ten sessions, resulting in ten successful authentications from a single code. This occurred because the HMAC computation is stateless, meaning the code is mathematically valid regardless of how many sessions submit it, and without server-side token tracking, nothing prevents reuse. When a used_tokens blacklist was enabled, only one session succeeded and the remaining nine were denied, confirming that explicit server-side state is required to enforce single-use behavior.

The push-based system demonstrated a structural advantage in this scenario. Because each login attempt generated a universally unique push identifier (UUID), cross-session confusion was architecturally impossible. Each push request could only be approved or denied within the context of its original session. All ten sessions were approved independently with zero collisions. This shows a fundamental design difference between the two methods. TOTP requires an explicit blacklist mechanism to achieve single-use behavior, while push-based MFA achieves it inherently through UUID assignments.

### *C. Test 3: State Loss and Server Restart*

The third experiment evaluated how each system behaved when server-side state was lost due to a restart. Authentication was initiated and completed successfully, after which the server was restarted mid-session to simulate an application crash, scheduled maintenance, or an attacker-induced denial-of-service event. Previously captured tokens and approval responses were then replayed against the restarted server.

**TABLE V**

*TOTP State Loss / Restart Test*

| Step | Action | Result |
|---|---|---|
| 1 | Login with token 170813 | Success |
| 2 | Replay before restart | Blocked |
| 3 | Server restarted | used_tokens wiped |
| 4 | Replay same token after restart | Accepted |

*Note. This table shows TOTP behavior before and after a server restart event.*

**TABLE VI**

*Push MFA State Loss / Restart Test*

| Step | Action | HTTP | Result |
|---|---|---|---|
| 1–3 | Login, approve, token issued | 200 | Active |
| 4 | Verify token before restart | 200 | Authorized |
| 5 | Server restarted | — | sessions{} wiped |
| 6 | Old token after restart | 401 | Unauthorized |

*Note. This table shows push MFA session behavior before and after a server restart event.*

The results reveal a shared vulnerability with distinct consequences for each system. In the TOTP implementation, replay prevention was enforced through an in-memory used_tokens list. Upon server restart, this list was cleared, and a previously captured token that had been correctly rejected before the restart was accepted without restriction afterward. This means any attacker capable of triggering a server restart, either through a denial-of-service attack or other means, could immediately replay a previously intercepted TOTP code to gain unauthorized access.

The push-based system exhibited a different failure mode. Session tokens were stored in an in-memory dictionary that was also wiped on restart. Unlike the TOTP case, this did not allow old tokens to be accepted. Instead, all active sessions were silently invalidated, returning a 401 Unauthorized response. While this prevents unauthorized access from replayed tokens, it introduces a denial-of-service condition in which legitimate users are forced to re-authenticate after every restart. Both outcomes highlight a common root cause: critical authentication state stored only in memory does not survive infrastructure events.

## V. CONCLUSION

This study looked at the gap between how secure MFA systems are supposed to be in theory and how they perform in practice. This project built and tested an MFA system with both TOTP and push-based authentication to see how well each one held up when testing freshness, session binding, and server-side state management.

What was found across all three tests was that the weaknesses uncovered weren't caused by the cryptography itself failing; they came from how the systems were implemented around those algorithms. When the validation windows were widened, token tracking was removed, or the server was restarted, both systems became vulnerable even though the underlying math was still correct.

One of the more interesting differences found was in how the two methods handle session binding. Push-based MFA has a natural advantage here because every login attempt gets its own unique ID, so there's no way to accidentally reuse an approval across sessions. TOTP doesn't have that built in. Without a server-side blacklist explicitly tracking used tokens, the same code can be accepted across multiple sessions at once.

Both systems also shared a common vulnerability tied to in-memory state storage. A server restart clears the used-token set that TOTP depends on for replay prevention and wipes the session dictionary that push-based authentication uses to validate active tokens. This means an attacker who can trigger a restart gains the ability to replay captured TOTP tokens, while all active push sessions are simultaneously terminated. Moving authentication state to persistent storage, such as a database, would directly address this weakness.

These findings support the broader conclusion that the security of an MFA system depends not only on the strength of its cryptographic algorithm, but also on the correct surrounding implementation. Proper window sizing, persistent state management, session binding, and single-use token enforcement are all necessary to make cryptographic guarantees hold in practice.

## REFERENCES

[1] J. Reynolds, N. Samarin, J. Barnes, T. Judd, J. Mason, M. Bailey, and S. Egelman, "Empirical measurement of systemic 2fa usability," in Proceedings of the 29th USENIX Conference on Security Symposium, ser. SEC'20, USA: USENIX Association, 2020, ISBN: 978-1-939133-17-5.

[2] J. Bonneau, C. Herley, P. C. v. Oorschot, and F. Stajano, "The quest to replace passwords: A framework for comparative evaluation of web authentication schemes," in 2012 IEEE Symposium on Security and Privacy, 2012, pp. 553–567. DOI: 10.1109/SP.2012.44.

[3] A. T. Mahdad, M. Jubur, and N. Saxena, "Breaking mobile notification-based authentication with concurrent attacks outside of mobile devices," in Proceedings of the 29th Annual International Conference on Mobile Computing and Networking, ser. ACM MobiCom '23, Madrid, Spain: Association for Computing Machinery, 2023, ISBN: 9781450399906. DOI: 10.1145/3570361.3613273. [Online]. Available: https://doi.org/10.1145/3570361.3613273.